# The Dark Side of Dark Mode

User behaviour rebound effects and consequences for digital energy consumption

Zachary Datson
BBC Research & Development
London, UK
zak.datson@bbc.co.uk

## ABSTRACT

User devices are the largest contributor to media related global emissions. For web content, dark mode has been widely recommended as an energy-saving measure for certain display types. However, the energy savings achieved by dark mode may be undermined by user behaviour. This pilot study investigates the unintended consequences of dark mode adoption, revealing a rebound effect wherein users may increase display brightness when interacting with dark-themed web pages. This behaviour may negate the potential energy savings that dark mode offers. Our findings suggest that the energy efficiency benefits of dark mode are not as straightforward as commonly believed for display energy, and the interplay between content colour-scheme and user behaviour must be carefully considered in sustainability guidelines and interventions.

## INTRODUCTION

Internet activity comes with a surprising environmental cost. Whilst a single webpage visit consumes a tiny amount of energy, the total estimated electricity footprint of the internet runs into the thousands of TWh per year, causing hundreds of megatons of CO2e emissions [1]. The energy consumption of user devices represents the most important fraction of this footprint, significantly outweighing the contribution of the shared internet infrastructure [1-3]. It is therefore vital for digital emission reduction efforts that we mitigate device energy consumption on the web [3].

A significant driver of energy consumption in user devices is the device display, as screen pixels use energy to emit light. Depending on the device and display type, at maximum brightness the display can account for nearly three quarters of total power consumption [4]. Two major display technologies dominate the current consumer electronics market – organic light-emitting diode (OLED) displays and liquid crystal displays (LCD). The principal difference between the two, from an energy perspective, is that LCDs employ a backlight, while in OLED displays, each pixel emits its own light independently. LCD energy consumption is therefore unaffected by the displayed content, whereas for OLED displays the two are directly related.

Choosing energy-conscious user interface (UI) colour palettes consequently has the potential to make real energy savings for devices with OLED displays. "Dark mode" – dark-themed content - is an example of this. Studies have found that switching to darker colour schemes can significantly reduce power consumption for devices with OLED displays [5-6]. Dark mode is therefore frequently recommended as a simple energy saving choice [7-8]. Blackle, a dark mode alternative to Google dreamed up in 2007, claims that, thanks to its black colour-scheme, over 10 million Watt-hours (Wh) have been saved since then [9]. Websites continue to implement dark mode today, making similar energy efficiency claims [10].

However, we cannot be certain that dark mode is truly saving energy without examining its usage in the wild, where user behaviour may cause unexpected rebound effects. One such effect is that users may increase their device brightness to compensate for the darker content, potentially negating any energy savings. This is especially relevant given that most devices still use LCDs, where power consumption may not be reduced by displaying darker colours. Without thorough consideration of user behaviour, endorsing dark mode as an effective low carbon measure remains a sceptical proposition. This study therefore investigates energy implications of user behaviour, comparing preferred brightness levels for dark mode and traditional light mode.

## METHOD

Experiments were conducted using a 2017 MacBook Pro (the device) running Ventura 13.6.4. The device used a 3.1 GHz Dual-Core Intel Core i5 processor, Intel Iris Plus Graphics and had 16GB memory and a 13.3-inch LCD display.

For the following tests, the BBC Sounds home page was used, which features a dark or light-mode version that can be set using system preferences (Figure 1). BBC Sounds power consumption for both dark and light versions was measured using a Tektronix PA1000 external power monitor, connected to the device via a breakout box. Tektronix PWRVIEW software was used to control the power monitor and log power data. To minimise unwanted activity, all background processes were terminated. Power consumption was measured for page load, followed by 15s of scrolling activity to simulate user interaction. This behaviour was automated using a custom Python tool implementing Selenium WebDriver (https://www.selenium.dev/).

To investigate user behaviour, participants (n=10) were seated 50cm from the test device displaying the BBC

Sounds home page and asked to alter the brightness until they were comfortable with the viewing conditions. Device brightness was reset to a baseline of 0 after every test. This test was repeated with the page in dark mode and light mode and in two different lighting environments: a dimly lit room and a brightly lit room. Each participant experienced all combinations of all conditions (dark mode in a dim room (DD), dark mode in a light room (DL), light mode in a dim room (LD), and light mode in a light room (LL)).

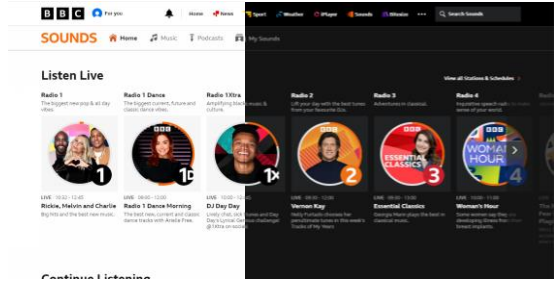

Figure 1: Light (left) and dark (right) mode for BBC Sounds home page.

## RESULTS

The power for a simulated visit to the BBC Sounds home page was recorded at 16 brightness levels, where level 1 was a dimmed screen and level 16 was full brightness. Measurements were repeated 5 times for each brightness level for both dark and light mode (Figure 2). There was no significant difference between measured page visit power for dark and light mode at any brightness level ($p>0.05$). Brightness level, however, significantly impacted power ($p<0.001$), and increasing display brightness resulted in increased power consumption.

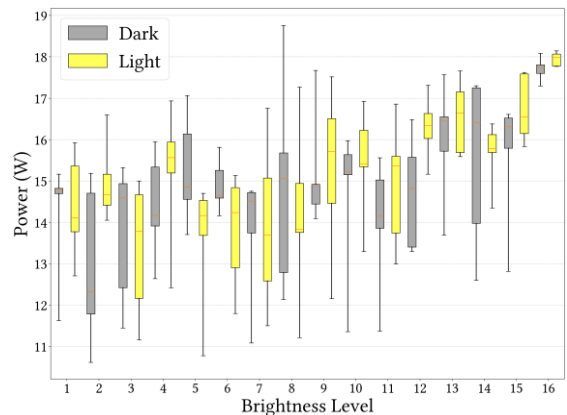

Figure 2: Average page visit power (W) for dark and light mode versions of the BBC sounds home page, with increasing display brightness.

For user tests, preferred brightness was recorded for each participant for each condition. Data was used to produce descriptive statistics (Table 1) and box-and-whisker plots for simple comparison of conditions (Figure 3).

Data analysis showed no significant difference between dimly and brightly lit environmental conditions ($p=0.13$). The colour scheme of displayed content however, strongly impacted preferred brightness level ($p<0.001$). Participants turned the test device brightness up higher for dark mode, irrespective of lighting environment.

|      | DD   | LD  | DL   | LL   |
|------|------|-----|------|------|
| mean | 12.5 | 9.6 | 12.7 | 10.7 |
| std  | 2.1  | 3.1 | 2.1  | 3.3  |
| min  | 8    | 5   | 9    | 5    |
| max  | 14   | 13  | 16   | 14   |

Table 1: Descriptive statistics for preferred brightness levels under different conditions: dark mode in a dim room (DD), dark mode in a light room (DL), light mode in a dim room (LD) and light mode in a light room (LL).

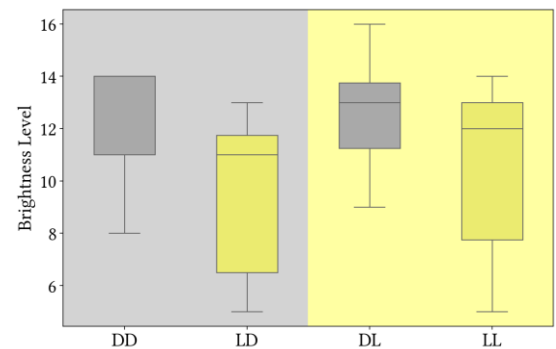

Figure 3: Distribution of brightness levels selected by participants under different conditions.

## DISCUSSION

Our findings show that, as expected for a device with an LCD, content colour scheme did not significantly impact energy consumption. Brightness was the only factor affecting display energy, and increasing display brightness caused increased energy consumption.

On the other hand, the results of our user study demonstrate that, at least for LCDs, content colour scheme does significantly influence users' preferred brightness levels. Participants preferred an increased display brightness when using dark mode, irrespective of the surrounding lighting conditions. As increasing brightness increased device energy use, this user behaviour rebound effect challenges the assumption that using dark mode saves energy. If our results can be validated across a larger sample, they could have substantial implications for digital emissions recommendations. Further investigation is also needed to determine whether the observed rebound effect applies to devices with OLED displays, and to quantify the energy trade-off.

## REFERENCES

[1] J. Malmodin, N. Lövehagen, P. Bergmark, and D. Lundén, 'ICT Sector Electricity Consumption and Greenhouse Gas Emissions – 2020 Outcome', Telecommunications Policy, Volume 48, Issue 3, 2024,. doi: 10.1016/j.telpol.2023.102701.

[2] A. S. G. Andrae and T. Edler, 'On Global Electricity Usage of Communication Technology: Trends to 2030', *Challenges*, vol. 6, no. 1, Art. no. 1, Jun. 2015, doi: 10.3390/challe6010117.

[3] M. Arora, I. McClenaghan, and L. Wozniak, 'Priorities for net-zero web services', *Nature Electronics*, vol. 7, no. 8, pp. 622–625, Aug. 2024, doi: 10.1038/s41928-024-01227-8.

[4] X. Chen, K. W. Nixon, H. Zhou, Y. Liu, and Y. Chen, 'FingerShadow: an OLED power optimization based on smartphone touch interactions', in *Proceedings of the 6th USENIX conference on Power-Aware Computing and Systems*, in HotPower'14. USA: USENIX Association, Oct. 2014, p. 6.

[5] D. Li, A. H. Tran, and W. G. J. Halfond, 'Making web applications more energy efficient for OLED smartphones', in *Proceedings of the 36th International Conference on Software Engineering*, Hyderabad India: ACM, May 2014, pp. 527–538. doi: 10.1145/2568225.2568321.

[6] P. Dash and Y. C. Hu, 'How much battery does dark mode save?: an accurate OLED display power profiler for modern smartphones', in *Proceedings of the 19th Annual International Conference on Mobile Systems, Applications, and Services*, Virtual Event Wisconsin: ACM, Jun. 2021, pp. 323–335. doi: 10.1145/3458864.3467682.

[7] T. Greenwood, 'The dark side of green web design', Wholegrain Digital. Accessed: Sep. 18, 2024. [Online]. Available: https://www.wholegraindigital.com/blog/dark-colour-web-design/

[8] 'Does dark mode win on sustainability and accessibility?', valtech.com. Accessed: Sep. 18, 2024. [Online]. Available: https://www.valtech.com/en-gb/blog/does-dark-mode-win-on-sustainability-and-accessibility/

[9] 'Blackle - Energy Saving Search'. Accessed: Sep. 19, 2024. [Online]. Available: http://www.blackle.com/

[10] 'EFRAG Website Launches an Energy-Saving Dark Mode Feature | EFRAG'. Accessed: Sep. 18, 2024. [Online]. Available: https://www.efrag.org/en/news-and-calendar/news/efrag-website-launches-an-energysaving-dark-mode-feature